\documentclass[a4paper,fleqn,usenatbib]{mnras}

\usepackage{newtxtext,newtxmath}

\usepackage[T1]{fontenc}
\usepackage{ae,aecompl}

\usepackage{graphicx}	
\usepackage{amsmath}	
\usepackage{xcolor}

\title[Predicting the flare productivity of emerging ARs]{On the possibility to probe the flare productivity of an active region on the early stage of emergence }

\author[A. S. Kutsenko et al.]{
Alexander S. Kutsenko,$^{1}$\thanks{E-mail: alex.s.kutsenko@gmail.com (ASK)}
Valentina I. Abramenko,$^{1}$
and Olga K. Kutsenko$^{1}$
\\
$^{1}$Crimean Astrophysical Observatory, p/o Nauchny, Crimea, 298409, Russia
}

\date{Accepted XXX. Received YYY; in original form ZZZ}

\pubyear{2020}

\begin{document}
\label{firstpage}
\pagerange{\pageref{firstpage}--\pageref{lastpage}}
\maketitle

\begin{abstract}
	
A long-standing problem is to predict the future flare productivity of an active region (AR) when it is on the stage of early emergence. The aim of this study is to probe two parameters of the photospheric magnetic field, both derived during the emergence phase of an AR, and to compare them with the flare productivity of the well-developed AR. The parameters are: (i) the index of the magnetic power spectrum (the slope of the spectrum) on the stage of emergence, and (ii) the flux emergence rate.  Analysis of 243 emerging ARs showed that the magnetic power index increases from values typical for quiet-Sun regions to the values typical for mature ARs within a day while the emergence proceeds for several days; frequently, after the increase, the value of the power index undulates around some mean value with fluctuations being several time less than the growth of the power index during the emergence onset. For a subset of 34 flare-productive ARs we found no correlation between the power spectrum index at the stage of emergence and flare index derived from the entire interval of the AR's presence on the disc. At the same time, the flux emergence rate correlates well with the flare index (the Pearson's correlation coefficient is 0.74). We conclude that high flux emergence rate is a necessary condition for an AR to produce strong flares in the future, and so, the flux emergence rate can be used to probe the future flare productivity of the AR.

\end{abstract}




\section{Introduction}
\label{sec:intro}

The solar physics community puts great efforts into development of approaches capable to predict solar activity on long term (month-years) time scales and solar flares on short term (hours-days) time scales. Solar flares -- intense explosive releases of electromagnetic radiation -- mostly take place in active regions (ARs). The energy that feeds solar flares is assumed to be stored in an AR in the form of the so-called `free' magnetic energy \citep[e.g.][]{Melrose1991, Schrijver2005}. This energy is accumulated in the corona due to non-potentiality of AR's magnetic field. It is supposed that the energy release takes place as a consequence of magnetic lines reconnections in the corona.

However, in the most of the cases, only data on photospheric magnetic field and continuum intensity are readily available for flare forecast. That is why the majority of flare prediction methods are based on the statistical correlation between AR parameters (usually derived from line-of-sight (LOS) or vector magnetograms measured at the photosphere level) and flare productivity. These parameters are the total vertical electric current \citep{Leka2007}, magnetic flux near strong-gradient polarity inversion lines \citep{Schrijver2007, Falconer2008}, fractal dimension measure of an AR \citep{McAteer2005}, multifractality spectrum \citep{Abramenko2005a}, the degree of non-neutralization of vertical electric current within an AR \citep{Kontogiannis2017}, and many others. More than a dozen of AR's parameters that were shown to be related to solar flare appearance \citep[e.g][and references therein]{Leka2003, Leka2007} are routinely calculated in real-time for Space-Weather Active Region Patches \citep[SHARPs,][]{Bobra2014} observed by the {\it Helioseismic and Magnetic Imager} on board the {\it Solar Dynamics Observatory} \citep[SDO/HMI,][]{Schou2012}. These data are arguably appropriate inputs for real-time solar flare predictions \citep[e.g.][]{Lee2018}.

Although a number of parameters derived from photospheric vector magnetic field exhibits mutual correlation with the flare productivity of an AR, \citet{Leka2007} argued that the state of the photospheric magnetic field at a given time provides limited information on whether the AR will produce a flare. It is worth noting that more recent approaches intend to add all the available data including chromospheric and coronal intensity maps in the forecast model \citep[e.g.][]{Jonas2018}. A variety of other methods such as machine learning \citep[e.g.][]{Bobra2015, Liu2017, Raboonik2017}, analyses of AR morphology \citep[e.g.][]{McIntosh1990, Bloomfield2012}, flare statistics \citep[e.g.][]{Wheatland2004}, radio emission fluxes \citep[e.g.][]{Tanaka1975, Bogod2018} are also widely used to solve the flare prediction problem.




The list of flare forecast approaches mentioned above is far from being a comprehensive one. \citet{Barnes2016} and \citet{Leka2019} compared the performance of currently operating solar flare prediction algorithms \citep[see the list of currently operating algorithms in table~1 in][]{Leka2019}. They concluded that the performance of different algorithms is quite comparable, i.e. there is no algorithm that performs much better than others. In the most of the cases the algorithms make their forecasts for a time interval of 24--48 h. The data on well-developed AR are usually required to calculate flare-related parameters used in most algorithms. For example, one can not estimate the vertical current along the polarity inversion line in emerging AR if this inversion line is not formed yet. Is there a property or parameter of an AR that can be used to predict AR's flare productivity in advance, say, as soon as the AR starts to emerge? Can we use this parameter to conclude that the AR is `born-bad', i.e., it is capable to produce strong flares in the future? Alternatively, can we reveal `quiet' ARs during early stages of their emergence? We suppose that at least two parameters that can be estimated during early stages of emergence of an AR -- the index of the magnetic power spectrum (power index) and the flux emergence rate -- might be predictors for the future flare productivity.

\citet{Abramenko2005} found a statistical relationship between the time-averaged power index -- the negative slope of the magnetic power spectrum -- of a well-developed AR and its flare productivity derived from the entire time interval when the AR was on the solar disc. High-resolution magnetograms acquired by the \textit{Michelson Doppler Interferometer} on board the \textit{Solar and Heliospheric Observatory} \citep[SOHO/MDI][]{Scherrer1995} were used to derive the power spectra and power indices for 16 ARs. \citet{Abramenko2005} showed that the power index tends to increase for ARs with higher flare productivity. Much more solid set of 217 ARs was considered by \citet{Abramenko2010}. The ARs were observed by SOHO/MDI between 1997 and 2006 and were divided into well-developed, emerging, decaying, and unipolar ARs subsets. The authors confirmed positive correlation between the power index and the flare productivity of an AR with Pearson's $\rho=0.57$. Unipolar spots exhibited the lowest values of both the flare index and the power index. No segregation between the subsets of emerging and decaying ARs was found.

Findings reported by \citet{Abramenko2005} and \citet{Abramenko2010} were recently supported by \citet{Mandage2016} who studied magnetic power spectra of 53 ARs observed by SDO/HMI. The ARs were divided into two subsets of flare-quiet and flare-productive (produced at least one M1 class flare) ARs. The analysis showed that 90\% of flare-productive ARs and only 12\% of flare-quiet ARs exhibited the power index greater than 2. The authors also argued that flare-quiet ARs may exhibit high day-to-day variations of the power index while the power index of flare-productive ARs remains stable at relatively high level days before and days after flare events. \citet{Mandage2016} concluded that the power index can be employed to forecast flares several days in advance.

Previous studies of the power index and flare productivity in ARs used data for predominantly well-developed ARs. Although 6 out of 16 ARs studied in \cite{Abramenko2005} were classified as emerging ARs, none of them was tracked since the very beginning of the emergence. Instead, SOHO/MDI started to observe an AR in a high-resolution mode as soon as the clear emergence of the magnetic flux was detected. Interestingly, the power index retrieved from the first provided magnetogram was quite close to those calculated from magnetograms of the AR in a well-developed state. In general, the power index seems to increase abruptly as soon as an AR starts to emerge \citep{Abramenko2005}. \citet{Hewett2008} studied the power index temporal evolution of NOAA AR 10488. They found the rapid change of the power index from 0.5 to 2.5 within a day \citep[see fig.~8 in][]{Hewett2008}. \citet{KutsenkoOK2019} analysed power index variations in four emerging ARs. All of them exhibited abrupt increase of the power index as soon as first imprints of magnetic flux appeared on the photosphere. An important point is that the observed value of the power index was typical for mature ARs and changed insignificantly during further evolution of the ARs.

One more AR parameter that can be estimated on the early stages of emerging is the flux emergence rate.  \citet{Canfield1975} suggested the appearance of solar flare to be often related to the interaction between old and new emerging magnetic fluxes. This was later supported by \citet{Fu2016} who concluded that the emergence of a new AR in the vicinity of pre-existed ones is associated with the increase of strong flares occurrence. Fast-emerging ARs may disturb a pre-existed magnetic configuration and initiate magnetic reconnections and eruptions \citep{Archontis2019}. Thus, \citet{Schrijver2009} summarized that the emergence of a highly-twisted magnetic flux bundle into pre-existed strong magnetic surroundings leads to strong flares. In addition, \citet{Golovko1986} and \citet{Ishkov2003} argued that high-rate magnetic flux emergence often results in the formation of structures with high magnetic field gradients that tend to erupt. These findings were recently supported by numerical simulations performed by \citet{Toriumi2017}. The authors concluded that complex magnetic $\delta$-spot structures producing the largest X-class flares are formed due to interaction of several emerging magnetic flux loops or due to emergence of a highly-twisted magnetic flux bundle.

The flux emergence rate is in general proportional to the peak magnetic flux of an AR \citep[e.g.][]{Otsuji2011, Norton2017, Kutsenko2019}. However, ARs with similar peak magnetic flux may exhibit significant variations of the flux emergence rate \citep{Abramenko2017}. Our analysis of time profiles of the flux emergence rate performed in \citet{Kutsenko2019} suggested that ARs often emerge monotonously at a constant rate. Hence, we suppose that the flux emergence rate can be estimated on the stage of early emergence.

The aim of this work is to find out whether the power index and the flux emergence rate derived during the early stages of the emergence can be used to predict the flare productivity of ARs in the future. To achieve the goal, we intend (i) to analyse temporal variations of the power index of a statistically significant set of emerging ARs, and (ii) to explore a relationship between each of our parameters 
and flare productivity.

\section{Data and Methods}

The data set for 423 emerging ARs prepared previously in \citet{Kutsenko2019} was utilized as a basis in this study. For each AR the data set contains a time series of patches extracted from full-disc LOS magnetograms acquired by SDO/HMI. The patches represent the magnetic field maps of isolated emerging ARs as well as magnetic field maps of the quiet Sun these ARs emerge from. Each AR was tracked as long as the centre of the patch was located within $60^{\circ}$ from the central meridian. Both the emergence onset and the peak total unsigned magnetic flux were observed within these longitudinal ranges. A number of parameters, such as the peak total unsigned magnetic flux, flux emergence rate, time of the emergence onset and of the emergence halt, are gathered in the data set \citep[see][for details]{Kutsenko2019}. The total unsigned magnetic flux was calculated as a sum of absolute magnetic flux density values in LOS magnetogram pixels multiplied by the projection-corrected pixel area. Only pixels with absolute magnetic flux density above 18 Mx~cm$^{-2}$ (triple noise level in 720-s LOS SDO/HMI magnetograms as measured by \citealt{Liu2012}) were used in the summation. Before summation, the $\mu$-correction was applied to each pixel in the magnetograms \citep{Leka2017}.

\begin{figure*}    
	\centerline{\includegraphics[]{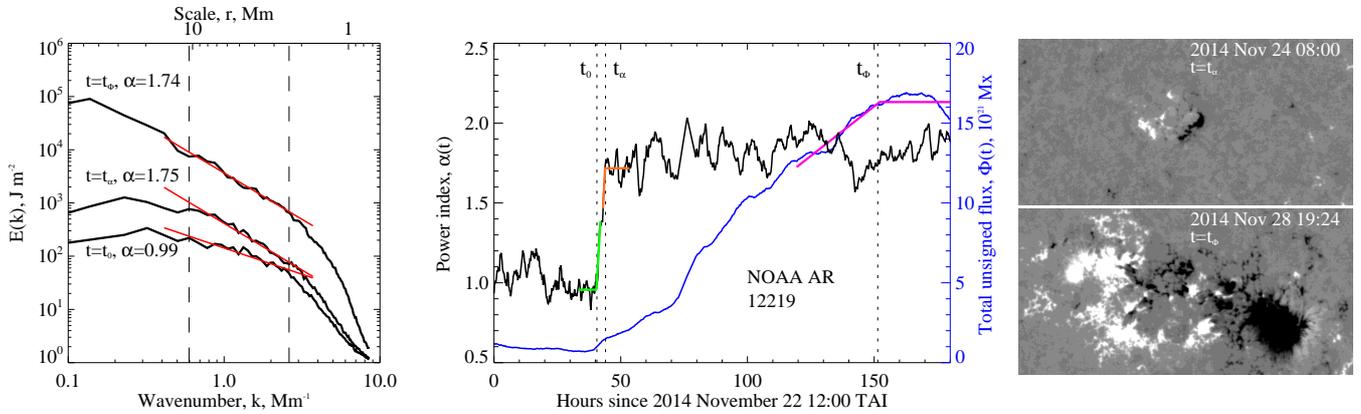}
	}
	\caption{ Left -- Magnetic power spectra of NOAA AR 12219 at times $t_{0}$, $t_{\alpha}$, and $t_{\Phi}$ (black curves). Red lines show the best linear fits to the spectra within the inertial range 2.4--10 Mm (enclosed between the dashed vertical segments). Middle -- Temporal variations of the power index $\alpha(t)$ (black curve) and of the total unsigned magnetic flux $\Phi(t)$ (blue curve) in NOAA AR 12219. Dotted vertical lines denote three times $t_{0}$, $t_{\alpha}$, and $t_{\Phi}$ determined from the fitting of $\alpha(t)$ and $\Phi(t)$ curves (see text). Examples of piece-wise linear fitting of $\alpha(t)$ and $\Phi(t)$ curves are shown by green, orange, and purple lines. Right -- SDO/HMI LOS magnetograms of NOAA AR 12219 acquired on 2014 November 24 at 08:00 TAI (corresponds to $t = t_{\alpha}$, top) and on 2014 November 28 at 19:24 TAI (corresponds to $t = t_{\Phi}$, bottom). The field-of-view is 200$\times$100 arcsec\textsuperscript{2}. The magnetograms are scaled from -500~Mx~cm\textsuperscript{-2} (black) to 500~Mx~cm\textsuperscript{-2} (white).
	}
	\label{fig1}
\end{figure*}

From the entire data set, we selected only ARs with assigned NOAA number and with the peak total unsigned magnetic flux between $0.6\times10^{21}$ and $26\times10^{21}$ Mx. The resulting subset consisting of 243 ARs was analysed in this study. The list of the ARs and the derived parameters described below are listed in Table~\ref{table1}. The full table is attached to this article as a supplementary material. For each NOAA AR, we calculated the magnetic power spectrum of the LOS magnetic field component. The procedure was developed in \citet{Abramenko2001} and was described in detail in \citet{Abramenko2005}. Briefly, a two-dimensional (2D) Fast Fourier Transform is applied to observed LOS magnetograms acquired at time $t$ yielding 2D spectra. An integration of the spectrum power over thin anulus between two concentric rings centred at the origin of the 2D-space of wavenumbers is performed to retrieve the 1D power spectrum $E(k, t)$, where $k=2\pi/r$ is the wavenumber and $r$ is the spatial scale in Mm. An example of the power spectrum evolution for NOAA AR 12219 is shown in the left panel of Fig.~\ref{fig1}. An important characteristic of a power spectrum is its slope. Throughout this paper we will refer to the negative slope of the power spectrum as the power index, $\alpha$. The power index depends on the turbulent regime, i.e., on the energy transfer mechanism in the medium where the magnetic field evolves \citep[e.g.][]{Kolmogorov1941, Biskamp1993}. Thus, the power index of quiet-Sun regions is of about 1.0 while that of AR ranges between 1.5 and 2.5. The power index is usually measured by fitting the power spectrum by a linear approximation within the inertial range. The inertial range is the range of wavenumbers where the spectrum is linear in a double logarithmic plot and a relationship $E(k, t) \propto k^{-\alpha}$ holds. The low-frequency (large-scale) limit of the inertial range relies on the scale where the dominant energy injection occurs (injection range). Thus, \cite{Mandage2016} argued that this limit may depend on the size of magnetic elements in the AR. The high-frequency (low-scale) limit is determined either by the dissipation, or, as in our case, by the (triple) spatial resolution of the telescope \citep{Abramenko2001}. Since we analyse both quiet-Sun and AR patches, in order to achieve the homogeneity of the measured power index we follow \citet{Abramenko2005} and define the inertial range to be between 2.4 and 10 Mm (Fig.~\ref{fig1}).

\begin{table*}
	\centering
	\caption{The list of ARs and the derived parameters. Note that one HARP may contain several ARs. The dates are in YYYY.MM.DD HH:MM format. The full table is available online as a supplementary material.}
	\label{table1}
	\begin{tabular}{ccccccccc} 
		\hline
		HARP  &Time of emerg.&Time of the power&Time of the peak& Saturated&Mean power&Peak total&Flux emergence&Flare\\
		number&onset (TAI)   &index saturation &magnetic flux&power&index, $\alpha_{mean}$&magnetic flux,&rate, $R_{av}$&index\\
		      &&(TAI)&(TAI)&index, $\alpha_{sat}$&&$\Phi_{max}$, $10^{21}$ Mx&$10^{20}$ Mx~h$^{-1}$&\\

		\hline		
		0014&2010.05.04 19:48&2010.05.05 02:24&2010.05.05 05:48&2.44$\pm$0.06&2.44$\pm$0.05&0.78$\pm$0.11&1.66$\pm$0.27&0.00\\
		0017&2010.05.05 20:12&2010.05.06 02:24&2010.05.07 12:24&1.88$\pm$0.05&1.42$\pm$0.27&0.80$\pm$0.13&0.43$\pm$0.08&0.00\\
		0026&2010.05.20 12:36&2010.05.20 22:00&2010.05.23 03:00&2.46$\pm$0.05&2.02$\pm$0.28&6.89$\pm$0.57&2.53$\pm$0.22&0.00\\
		0043&2010.05.30 21:48&2010.05.31 00:24&2010.06.03 06:24&1.54$\pm$0.16&1.66$\pm$0.29&7.55$\pm$0.61&2.18$\pm$0.18&0.00\\
		0049&2010.06.07 13:00&2010.06.07 22:24&2010.06.09 23:00&1.57$\pm$0.03&1.45$\pm$0.19&1.37$\pm$0.20&0.54$\pm$0.08&2.07\\
		0051&2010.06.09 17:48&2010.06.10 01:00&2010.06.11 11:36&2.21$\pm$0.02&2.15$\pm$0.20&3.64$\pm$0.37&1.99$\pm$0.21&0.24\\
		0054&2010.06.11 02:48&2010.06.11 21:12&2010.06.12 07:48&2.15$\pm$0.00&2.06$\pm$0.09&5.85$\pm$0.64&4.47$\pm$0.53&2.84\\
		0083&2010.07.03 21:12&2010.07.04 20:48&2010.07.05 09:24&2.18$\pm$0.07&2.15$\pm$0.09&1.78$\pm$0.17&1.14$\pm$0.11&0.00\\
		0116&2010.08.08 13:12&2010.08.08 16:24&2010.08.10 09:48&1.89$\pm$0.02&1.77$\pm$0.08&5.17$\pm$0.59&2.52$\pm$0.31&0.00\\
		0131&2010.08.10 19:12&2010.08.11 03:24&2010.08.11 18:24&2.46$\pm$0.03&2.47$\pm$0.13&2.68$\pm$0.37&2.17$\pm$0.36&0.00\\	
		\hline
		
	\end{tabular}
\end{table*}

Time variations of the power index, $\alpha(t)$, and of the total unsigned magnetic flux, $\Phi(t)$, in NOAA AR 12219 are shown in the middle panel of Fig.~\ref{fig1}. An abrupt increase of the power index is clearly seen just after the emergence onset while the total unsigned magnetic flux exhibits relatively gradual growth \citep[cf. fig.~3 in][]{KutsenkoOK2019}. Magnetic power spectra for three times labelled $t_0$, $t_\alpha$, and $t_\Phi$ are shown in the left panel of Fig.~\ref{fig1}. The time $t_0$ corresponds to the emergence onset when both power index and magnetic flux start to increase. The spectrum at $t_0$ has a shallow slope of about -1, which is typical for quiet Sun. As the emergence proceeds an injection of magnetic energy occurs at all spatial scales. However, the energy injection on larger scales prevails resulting in the steepening of the spectrum. The power index saturates at time $t_\alpha$ and fluctuates insignificantly around some mean value during further emergence (the fluctuations are several times less in magnitude than the total increase of the power index during the emergence). Interestingly, by this time only first imprints of emerging AR are seen in the magnetogram (Fig.~\ref{fig1}, right top panel). The energy growth proceeds on all spatial scales due to emergence of large magnetic elements and/or simultaneous emergence and merging of smaller magnetic features \citep{Hewett2008, KutsenkoOK2019}. Although the total energy increases (the spectrum moves upward, see the left panel in Fig.~\ref{fig1}), the slope of the spectrum changes insignificantly. The upper spectrum in the left panel of Fig.~\ref{fig1} corresponds to the time of the peak magnetic flux, $t_{\Phi}$.

Using the data on time variations of the power index and of the total magnetic flux for each AR, we calculated a set of parameters:

\begin{enumerate}
\item The time of the emergence onset, $t_0$, and the time of the power index saturation, $t_{\alpha}$. These times were determined by two fittings of the $\alpha(t)$ curve with two-segment piece-wise linear functions (shown in green and orange lines in the middle panel of Fig.~\ref{fig1}). For each fitting, one segment was required to be horizontal while the other could have arbitrary slope. The desired time point was defined as an intersection of the horizontal and inclined segments.

\item The time of the peak total unsigned magnetic flux, $t_{\Phi}$ was similarly determined by fitting the $\Phi(t)$ curve with a two-segment piece-wise linear function (marked with purple lines in the middle panel of Fig.~\ref{fig1}). 

\item Saturated power index $\alpha_{sat}$ was defined as the mean value of the power index over the $[t_{\alpha};t_{\alpha} + 1 h]$ interval. Recall that the inertial range was set to 2.7--10 Mm. Consequently, only magnetic elements larger than 2.7 Mm may contribute to the slope of the power spectrum. The characteristic lifetime of such elements is of order of hours. Therefore, we suppose that the calculation of the saturated power index as the mean value of the power index within 1-hour interval is reasonable. The uncertainty in the saturated power index, $\sigma _{\alpha_{sat}}$, was estimated as the standard deviation of the $\alpha_{sat}$ during this 1-hour interval $[t_{\alpha};t_{\alpha} + 1 h]$. Note that the small-scale noise in the magnetograms affects exclusively high-frequency tail of the power spectrum (beyond the inertial range) and does not influence the slope of the power spectrum.

\item  Mean power index value $\alpha_{mean} = \langle \alpha(t) \rangle \vert_{t_{\alpha}}^{t_{\Phi}}$ Similar to previous item, the uncertainty in the mean power index, $\sigma _{\alpha_{mean}}$, was estimated as the standard deviation of $\alpha_{mean}$ over the $[t_{\alpha};t_{\Phi}]$ interval. 
\item Duration of the power index increase $\Delta t_{\alpha} = t_{\alpha} - t_{0}$.
\item Duration of the magnetic flux emergence $\Delta t_{\Phi} = t_{\Phi} - t_{0}$.
\item Peak total unsigned magnetic flux $\Phi_{max} = \Phi(t = t_{\Phi})$. The error in $\Phi_{max}$ was calculated as $\sigma _{\Phi_{max}} = \sum_{n} 3 \sigma_{B} s$, where $3\sigma_{B}$=18 Mx~cm$^{-2}$ is the triple noise level in the SDO/HMI magnetograms \citep{Liu2012}, $n$ is the number of pixels used to calculate the total unsigned flux at $t = t_{\Phi}$, $s$ is a projection-corrected area of one pixel.
\item Flux emergence rate $R_{av} = (\Phi_{max} - \Phi(t_{0}))/\Delta t_{\Phi}$. The uncertainty of the flux emergence rate is proportional to the relative uncertainty of the peak magnetic flux and was estimated as $\sigma _{R_{av}} = \frac{\sigma _{\Phi_{max}}}{(\Phi_{max} - \Phi(t_{0}))} \cdot R_{av}$.

\item To evaluate the flare productivity of an AR, we, following \citet{Abramenko2005}, calculated its flare index (FI). FI is a normalized measure of X-ray radiation emitted by an AR:
\begin{equation}
\mathrm{FI} = (100S^{(X)} + 10S^{(M)} + 1.0S^{(C)} + 0.1S^{(B)})/\tau,
\end{equation}
where $S^{(j)} = \sum_{i=1}^{N_{j}} I_{i}^{j}$ is the sum of \textit{GOES} peak intensities of a certain flare class ($X$, $M$, $C$, $B$), $N_{j}$ is the number of flares of a certain class, and $\tau$ is the total time interval of AR observation in days \citep{Abramenko2005}. The data on FI was provided by the CrAO Solar Application web resource available at \verb|http://solar.dev.argh.team|. The website aggregates data on ARs and flares from different web sources. It provides information on AR's first and last dates of observation, on the class and power of the most intense flare produced by the AR, on the duration of the AR presence on the disc, and on the AR's flare index. The noise level of GOES fluxes is of about $2\times 10^{-8}$ W~m$^{-2}$ \citep{Aschwanden2012} that is two orders of magnitude less than C1.0 class flare. Consequently, we suppose that the uncertainties in the flare index measurements are negligible.

\end{enumerate}

\section{Results}

In all, among 243 emerging ARs analysed in this study, only 34 of them exhibited the flare index FI equal to or exceeding unity (at least one C1 class flare per day). We will refer to this 34 ARs as a set of flaring ARs through the rest of the paper.
Such a low fraction of flare-productive ARs in the set can be explained, first, by low activity of Solar Cycle 24, and, second, by strong selection rules: both the emergence onset and the peak magnetic flux must be observed within $\pm60^{\circ}$ from the central meridian. In addition, a majority of newly-emerged ARs were quite weak to produce strong flares or the flaring might start after the ARs moved to the far-side of the Sun.

\subsection{Temporal evolution of the power index}

\begin{figure}    
	\centerline{\includegraphics[width=\linewidth]{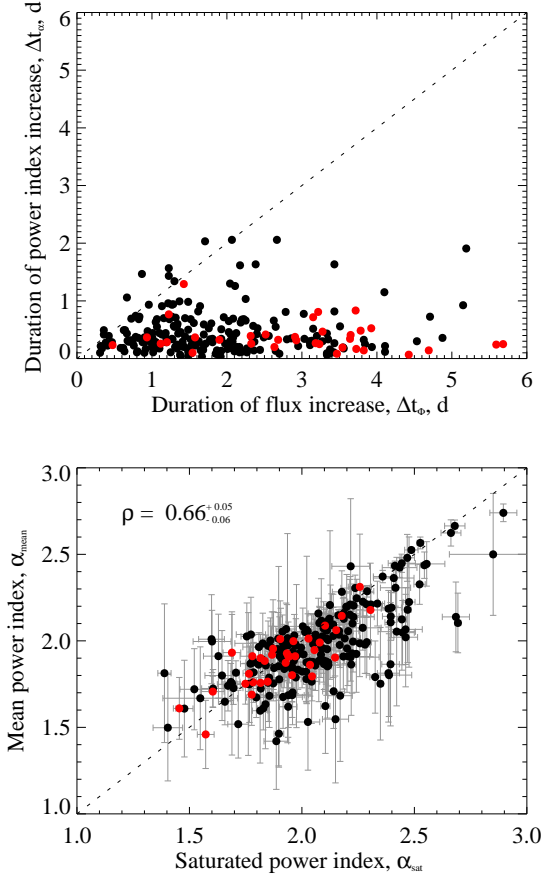}
	}
	\caption{
		Top -- Duration of the power index increase versus duration of the total unsigned magnetic flux increase for a set of 243 emerging ARs.  Bottom -- the mean power index $\alpha_{mean}$ versus the saturated power index $\alpha_{sat}$; error bars show standard deviation of the power indices (see  text).  Pearson's correlation coefficient $\rho$ is shown. The confidence interval is shown for 95\% confidence level. On both panels, red circles label the 34 flaring ARs, and dotted lines show the bisectors.
	}
	\label{fig2}
\end{figure}

Duration of the magnetic flux emergence, $\Delta t_{\Phi}$, versus duration of the power index increase, $\Delta t_{\alpha}$, for the entire data set is shown in the upper panel of Fig.~\ref{fig2}. The plot shows that in the most of the cases $\Delta t_{\alpha}$ does not exceed one day (11 h on average) while the flux emergence may last for several days. Consequently, the value of saturated power index can be estimated during initial stages of AR's emergence. A dotted diagonal line is plotted to emphasize that for the majority of the cases, the power index increase is faster than the increase of the total unsigned flux (note also equal axis scales). Red circles in Fig.~\ref{fig2} indicate the 34 flaring ARs. As one can see, there is no clear difference between the distributions of flare-productive and flare-quiet ARs.

The lower panel of Fig.~\ref{fig2} shows the saturated power index $\alpha_{sat}$ versus the mean power index $\alpha_{mean}$. Error bars in the plot are standard deviations of the power indices described in the previous section. The Pearson's linear correlation coefficient of the distribution equals 0.66 while the confidence interval is shown for 95\% confidence level. The corruption of the correlation coefficient due to uncertainties in the measured data was taken into account using the technique described in \citet{Saccenti2020}. Most of the data points are concentrated along the diagonal of the plot suggesting that on average the saturated power index (measured during the early stages of emergence) is a good estimator of the power index observed during the entire duration of emergence.

\subsection{Power index and flare productivity of newly-emerged ARs}

\begin{figure}    
	\centerline{\includegraphics[width=\linewidth]{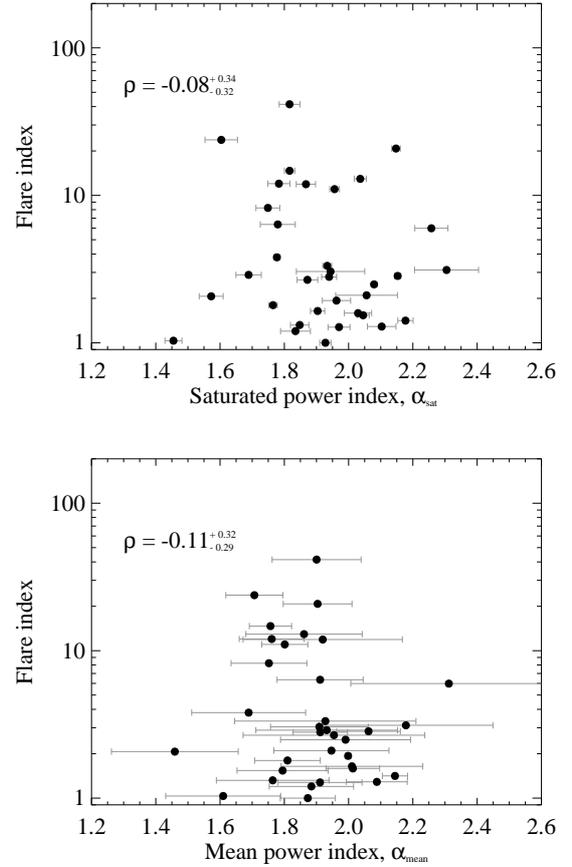}
	}
	\caption{
		Flare index of the 34 flaring ARs versus saturated power index $\alpha_{sat}$ (top) and mean power index $\alpha_{mean}$ (bottom). Pearson's correlation coefficients $\rho$ are shown. The confidence intervals are shown for 95\% confidence level.
	}
	\label{fig3}
\end{figure}

The relationships between the flare index, FI, and power index values $\alpha_{sat}$ and $\alpha_{mean}$ are shown in the top and bottom panels of Fig.~\ref{fig3}, respectively. In these plots we used data on flaring ARs with FI$\ge 1$. The plots show the absence of correlation between flare and power indices. This result might be caused by a specific choice of ARs in this work and is discussed in detail in Section \ref{sec:conclusions}.

\subsection{Flux emergence rate and flare productivity of newly-emerged ARs}

\begin{figure}    
	\centerline{\includegraphics[width=\linewidth]{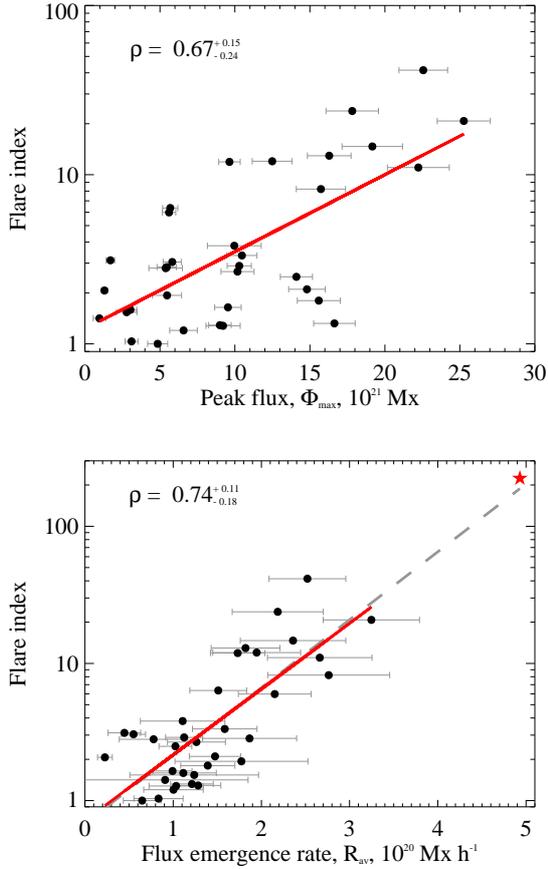}
	}
	\caption{
		The flare index versus the peak total unsigned magnetic flux $\Phi_{max}$ (top) and flux emergence rate $R_{av}$ (bottom) for the 34 flaring ARs. Pearson's correlation coefficients $\rho$ are shown. The confidence intervals are shown for 95\% confidence level. Red lines show the best linear fit to the data points. The red star in the lower panel marks a data point for NOAA AR 12673. A grey dashed line in the lower panel shows the best linear fit for the joint set of 35 ARs, including the NOAA AR 12673 (in this case $\rho$ increases up to 0.82).
	}
	\label{fig4}
\end{figure}

The scatter plots of the flare index FI versus the peak magnetic flux $\Phi_{max}$ and FI versus the flux emergence rate $R_{av}$ are shown in the upper and lower panels of Fig.~\ref{fig4}, respectively. The flare index exhibits a good positive correlation with the peak total unsigned flux with Pearson's correlation coefficient $\rho=0.67$ that is in a good agreement with previous results \citep[e.g.][]{Leka2007}. The correlation between the flux emergence rate and the flare index is even better. The Pearson's correlation coefficient $\rho$ reaches 0.74 for the data set of the 34 flaring ARs. Possible implications of this finding are discussed in the next section \ref{sec:conclusions}. Note, however, that there are ARs in our set of 243 ARs that exhibited high flux emergence rate but produced no significant flares.

An interesting example of fast-emerging and flare-productive AR is NOAA AR 12673 that was the most flare-productive AR in Solar Cycle 24. The AR produced a series of M- and X-class flares including the strongest X9.3 flare on 2017 September 06 \citep[e.g.][]{Wang2018}. The AR appeared on the Eastern limb of the solar disc as a decaying unipolar spot. During the passage across the disc, an intense emergence of a new magnetic flux started to the East from the pre-existed sunspot \citep[see][for a morphological analysis of AR's evolution]{Yang2017}. \citet{Sun2017} found that the averaged flux emergence rate was $4.93\times 10^{20}$~Mx~h\textsuperscript{-1} that was higher than ever reported previously. Although NOAA AR 12673 did not formally meet our requirements on emerging ARs (this AR did not appear amidst quiet Sun), we plotted $R_{av}$ and FI for this one on the bottom panel of Fig.~\ref{fig4} by a red star symbol. The position of the star agrees very well with the relationship between the averaged flux emergence rate and the magnitude of the flare index as derived from the set of the 34 flaring ARs. Adding this data point to the set of flaring ARs increases the linear correlation coefficient between $R_{av}$ and $log(FI)$ and yields Pearson's correlation coefficient $\rho = 0.82$. This experiment with NOAA AR 12673 allows us to suggest that a general tendency of the relationship between the flux emergence rate and the flare index is captured correctly.

\section{Conclusions and Discussion}
\label{sec:conclusions}

We employed SDO/HMI data for 243 emerging ARs to carry out a statistical analysis of temporal variations of the magnetic spectrum power index during the emergence of an AR. We found that the power index starts to increase abruptly as soon as the first imprints of magnetic flux appear on the photosphere. The duration of the power index growth from values typical for quiet-Sun to higher values typical for ARs usually does not exceed 1 day. This interval is several times shorter than the typical duration of magnetic flux emergence lasting up to 6 days for the largest AR observed in this study.
As the emergence proceeds, the power index usually stabilizes at values typically observed in well-developed ARs and exhibits just slight fluctuations (with magnitude being several times less than the power index increase) throughout the whole process of emergence. Consequently, in most cases we can evaluate the averaged value of the power index of a well-developed AR in advance if we observe the AR since the onset of emergence.

We found no correlation between the power index of emerging ARs and their future flare productivity. The reason for this negative result might be our restrict selection rules, and therefore, a scanty data set, only 34 cases. Besides, the very low activity of Solar Cycle 24 was the cause for the absence of strong flaring ARs in our data set. 
Note that \citet{Abramenko2005} analysed strong well-developed ARs of Solar Cycle 23 with a wide range of flare index ranging from 0 to 130. In the present data set, the highest flare index magnitude is 41 for NOAA AR 11158, only two ARs exhibit the flare index exceeding 20 and 5 ARs have flare index exceeding 10. In fig.~8 in \citet{Abramenko2005} there is a large scatter of power indices values (from 1.8 to 2.1) for the range of flare index 10--50 that is in accordance with our results. Seemingly, our restrictions on emerging ARs do not allow us to include more flare productive ARs.
Indeed, we required both the emergence onset and halt to be observed within 60$^{\circ}$ from the central meridian, i.e. the total interval of the AR tracking was limited to approximately 9 days. In that case, the process of emergence of strong ARs capable to produce strong flares might not be captured and/or the flaring might start when the AR rotated to the far-side of the Sun.
Summarizing, we conclude that the power index of the magnetic spectrum, as a predictive parameter, deserves further investigations based on data for other solar cycles, more strong than the analysed here Solar Cycle 24. 

An interesting result that can have implications for solar flare forecasting is the relationship between the flare index and the flux emergence rate of ARs. For the 34 flare-productive ARs (FI $\ge 1$) the logarithm of the flare index was in a good linear correlation with the flux emergence rate (Pearson's correlation coefficient is 0.74). Note that in contrast to previous researches mentioned in section \ref{sec:intro}, in the present study we considered only isolated ARs emerging amidst quiet-Sun areas. Our results suggest that fast-emerging ARs tend to flare even if they emerge into areas with no pre-existed magnetic flux.

In our data set of 243 ARs there were a lot of rather small ARs that exhibited high value of flux emergence rate but produced no significant flares. We argue that the high  flux emergence rate is a necessary but not sufficient condition for an ARs to produce strong flares. At the same time, one should keep in mind that at least a part of ARs with high flux emergence rate start flaring at the far-side of the Sun after they rotate off the western limb.

The flux emergence rate measurements might be employed to make early forecast of AR's flaring activity. First, the flux emergence rate can be continuously re-evaluated as the emergence proceeds, and therefore, the future flare productivity might be continuously evaluated more precisely. Second, the flux emergence rate is an easy-to-measure quantity that can be evaluated from, say, ultraviolet maps of an AR \citep{Schrijver1987}. This is an advantage of $R_{av}$ for solar flare forecasting compared to other quantities since emergence rate measurement does not require sophisticated magnetic field data. Thus, \citet{Ugarte-Urra2015} used 304~\AA\ ultraviolet observations provided by the {\it Solar Terrestrial Relations Observatory} \citep[STEREO,][]{Kaiser2008, Howard2008} to analyse the total unsigned magnetic flux of ARs emerging at the far-side of the Sun. Consequently, prospective instruments orbiting the Sun similar to STEREO satellites can provide information on magnetic flux variations of all ARs emerging on the solar surface. 

\section*{Data Availability}
The data underlying this article are available in the article and in its online supplementary material.

\section*{Acknowledgements}
We are grateful to the anonymous referee whose comments helped us to improve the paper. SDO is a mission for NASA's Living With a Star (LWS) programme. The SDO/HMI data were provided by the Joint Science Operation Center (JSOC). Analysis of the relationship between the power index and flare productivity of active regions was supported by the Russian Science Foundation (Project 19-72-00027).







\bsp	
\label{lastpage}
\end{document}